\documentstyle[preprint,epsfig,aps]{revtex}
\begin{document}
\draft
\tightenlines

\title{\bf{Statistics of Red Sites  on Elastic and Full Backbone}}

\author{Parongama Sen}

\address{Institute of Theoretical Physics,
University of Cologne,}
\address {Zulpicher Strasse 77,
50937  Cologne, Germany, }
\address {e-mail paro@thp.uni-koeln.de}
\maketitle
\narrowtext
\begin{abstract}

We investigate the number of red sites  on the elastic and
real backbone when right at the percolation threshold a spanning cluster
 exists between two sites at  opposite faces of the lattice and 
found that it scales 
in the same way as in the case of percolation between two plates.
 We also find out  that the number of common  
red sites scales similarly for  both kinds of backbones for percolation
between  pairs of sites on opposite faces of the 
lattice. 
Our statistics for several quantities show that
the the exponent for the elastic backbone approaches  the one of the 
full backbone as 
dimensionality is increased.

\end{abstract}

\vskip 2cm
\pagebreak

  Studies on cluster   structure dates back to more than two decades.
 Initially, de Gennes [1] 
and Skal and Shklovskii [2] independently postulated that 
the backbone at the percolation threshold consists of a network of 
nodes connected by 
effectively one-dimensional links. 
This picture turns out to be correct for dimensions greater than six, but
is too simplistic for low dimensions. Consequently, the  nodes, links and 
 blobs picture was postulated by Stanley [3] where the nodes are connected
 by one dimensional links
which  are often separated by multiconnected pieces or blobs.
In the backbone (without the dangling ends) there exist two kinds of bonds, 
the red (blue) bonds,  which, if cut, separate (do not separate) the 
infinite cluster in two separate clusters. 
However, this definition is perfect only as long as there is only
one spanning cluster; if there are more than one 
spanning cluster (which has a finite probability [4], especially in higher
dimensions), an infinite cluster
 still exists 
if  red bonds of 
another spanning cluster are  cut.
Hence, in general, we define, a red bond is that which when absent, cuts 
 one spanning cluster in the lattice which may still have 
other spanning clusters; the traditional definition allows for configuration  
without any red bonds [5].
It was proved by Coniglio [6] that the number of red bonds scales 
as $(p - p_c)^{-1}$, 
which was supported by Pike and Stanley [7] who showed numerically for 
two dimensions that the number
of red bonds  indeed follows 
the above scaling. 

  There  can be two different ways of realising  a 
percolating lattice. Either 
one can set all the sites of the top and bottom surfaces of the lattice to be
conducting (percolation between two conducting plates which we 
abbreviate as LLP - line to line percolation) or one can have a 
percolating
path between a pair of sites on the opposite faces of the lattice (SSP or
site to site percolation). In the latter, not all sites are occupied
in the top and bottom surfaces.  A percolating path may be 
regarded as occurring between site $a$ on the top surface to site $b$ on the
bottom surface, both belonging to the infinite cluster.   
Therefore,  there can be a number of paths connecting 
the top and bottom of the lattice depending on which two points 
are chosen at the top and bottom surfaces respectively. 

Skal [8] argues that there 
is another class of special
bonds, which apparently are similar to red bonds 
but follows a totally different scaling law  $(p-p_c)^a$ where
$a $ is positive.  
These special bonds are the ones through which every path connecting 
the opposite faces of the lattice must pass (for site to
site percolation). 
 One can obtain
several  paths  between the  
opposite faces of a lattice, each with a number of red bonds of its 
own, but there will be some which 
cannot be avoided by  any path. At $p_c$, according to [8], there is only 
one such 
bond and its number increases away from $p_c$.

  Herrmann et al [9] defined another fractal object which
they called the elastic backbone - a cluster of the 
sites that lie on the union of the shortest paths between two 
points.
  The number of cutting bonds or red bonds 
for the elastic backbone  at $p_c$ 
found by them follows  a different behaviour compared with that on the 
full backbone for two and three dimensions.
 The elastic backbone comprises of links and multiconnected paths where 
each path has the same length.  The full backbone can be obtained by
growing it from the elastic backbone.

In this paper, we have primarily investigated whether the 
scaling laws for the red sites  (either on the real or elastic
backbone) are affected by the above two boundary conditions in the
sense, whether it depends on the fact that the percolating 
path is between two sites or two plates.
 Throughout the simulation, we have considered
site percolation, and it is expected that red sites and bonds behave
in the same way.  

We have also shown that the quantities like the red sites, 
shortest path and mass of the elastic backbone (but not the mass 
of the full backbone) have essentially
the same exponent (= $ D_{min}$ where $D_{min} $ is the shortest
path exponent at $p_c$ [10]) in all dimensions. The exponent $D_{min}$ for the
shortest path on the elastic backbone is as expected, as by definition, the
elastic backbone is the shortest path. However, the result that the 
mass and the number of red sites on the elastic backbone also have the same 
exponent is not obvious. This indicates  
that the elastic 
backbone, even at
lower dimensions, corresponds  more than the full backbone to the description
of the node and links picture. 
 Hence it is not surprising, that 
the elastic backbone exponents and the backbone ones 
come closer as the dimensionality is increased.

Next we  investigate the scaling law for the number of 
common red sites  which lie in all the different connecting 
paths in SSP on both kinds of backbones.
If a common site is taken away,   
 there can be no spanning cluster at all in case we 
had before one spanning cluster only. These
are related to what Skal [8] calls the special bonds. In case there are
several spanning clusters, obviously there cannot be any common
red bond between them. We, however, study  an average picture 
here, rather than a distribution of the common red sites. Apparently,
it should be zero when there are more than one cluster as mentioned
earlier or
in the extremely unlikely cases where the different paths connecting
the top and bottom surfaces do not share a red site  at all. 

We have also detected the presence of what can be called a "breakthrough"
site.  Starting  from a rough estimate of $p_c$, say $p_0$, we can find out 
whether a cluster percolates from top to bottom, when sites are present with
 the probability $p_0$.  The probability $p_0$ is subsequently decreased 
(increased)  if there exists (does not exist) a percolating cluster
in the same manner as in [8].  It is observed that very close to $p_c$,
where this process is continued, 
 there can occur a spanning cluster depending on whether
or not   {\it one } 
particular site is present. This  was found to be true for 
all dimensions and sizes.
Defined in this way, there is exactly one  "breakthrough" site for each
percolating cluster which appears 
or disappears to enable percolation and is in some sense 
reminiscent of the "Starry sky model" as in [8].

We have carried out simulations at the percolation
threshold in dimensions $d$ = 2 to 5 for lattices
of size $L^d$ where $L$ is the linear dimension. In case of SSP,
randomly some connected paths were chosen 
to find out the number
of common red sites. 
The results should not depend on the number of random paths chosen as long 
as it is greater than 1. To be on the safe side, we have taken typically
10 to 30 such paths. It has been checked that indeed the results are
independent of the number of paths chosen in two dimensions.

We have generated the elastic backbone by a forward and backward
burning process [7]. In the case of site to site percolation, not all
the sites of the top and bottom surfaces are occupied. One can choose any site 
on the bottom surface belonging to the spanning cluster and start the 
backward burning. Backward burning essentially traces the shortest path to 
the upper surface. Therefore we have control in selecting one of the sites
through which site to site percolation is occurring, the other is 
fixed by the fact that it has to be the shortest path.  The backbone has been 
obtained by adding those blobs 
in the incipient cluster which are joined to the elastic backbone 
at two different points.
  The path length, mass, red sites
and  common red sites for elastic backbones in different dimensions are shown
 in Fig. 1 for   SSP and in Fig. 2 for LLP.
 The numerical values are different in the two cases but
they all have the same slope in these log-log plots. However, in SSP, the 
path length and mass almost converge indicating the presence of fewer blobs 
in the elastic backbone.
That this is not the case in the second case is obvious, there being 
a larger mass contribution due to the fully occupied top and
bottom surfaces. 
Here also, the finite size effects are more 
prominent due to the same reason. 

We have  been able to study the number of red sites and the common 
red sites  in the full backbone only for very small sizes due to the 
large amount of time involved to do this. However, the data still  
indicate that the  number of red sites and common red sites
 follow the Coniglio law for dimensions 2 to 4.
The LLP data is expected to be more inaccurate due to the 
small sizes and this is indeed the case as seen in Fig. 3 (especially
in case of $d$ = 4).

Finally, we investigate the ratio of the elastic backbone mass ($M_e$) to 
the real backbone mass  ($M_b$) and find that $ M_e \sim M^{\alpha}_b$.
We found that  $\alpha $ approaches unity as the  dimensionality is
increased as shown in Fig. 4.
This again supports the idea that the properties of the elastic 
backbone and the full backbone converge in higher dimensions.

Hence we conclude that there are three independent exponent
for red sites 
(a) $D_{min}$,  for red sites and common red sites on elastic backbone.
(b)   $1/\nu$, for red sites and common red sites on full backbone.
(c)  0, for the breakthrough  site which decides whether there
will be percolation or not when the concentration is varied.

\vskip 1cm

 This work is supported by  SFB 341.  The author expresses sincere
gratitude to Dietrich Stauffer, Hans J. Herrmann, Asya S. Skal and 
Michael Aizenman for suggestions,  discussions and 
encouragement.

\pagebreak

\begin{figure}

\caption {The number of red sites ($\diamond$), shortest path ($+$) and
 mass ($\Box$) of the 
elastic backbone are shown for percolation between two plates in (a) two
(b)three (c) four and (d) five dimensions. 
The dotted line has a slope 
$D_{min}$ =   (a) 1.13, (b) 1.34, (c) 1.50 and (d) 1.80. }

\end{figure}
\begin{figure}
\caption {The number of red sites ($\diamond$), shortest path ($+$), 
mass ($\Box$) and 
common red sites ($\times$)  of the 
elastic backbone are shown for percolation between two sites in (a) two
(b)three (c) four and (d) five dimensions.
The dashed line has a slope 
$D_{min}$ =   (a) 1.13, (b) 1.34, (c) 1.50 and (d) 1.80. }
\end{figure}
\begin{figure}
\caption {The number of backbone red sites ($\diamond$ for LLP, $+$ for SSP)  and 
common red sites for SSP ($\Box$) 
are shown for (a) 2 (b) 3 and (c) 4 dimensions. 
The dotted line has a slope 
$1/\nu$ =   (a) 0.75, (b) 1.13  and (c) 1.47. }
\end{figure}
\begin{figure}
\caption {The mass of the full backbone vs. the mass of the 
elastic backbone are shown for 2 ($\diamond$), 3($+$), 4($\Box$) and 
5 ($\times$)  dimensions. The dashed line has slope 1.}
\end{figure}
\psfig {file = 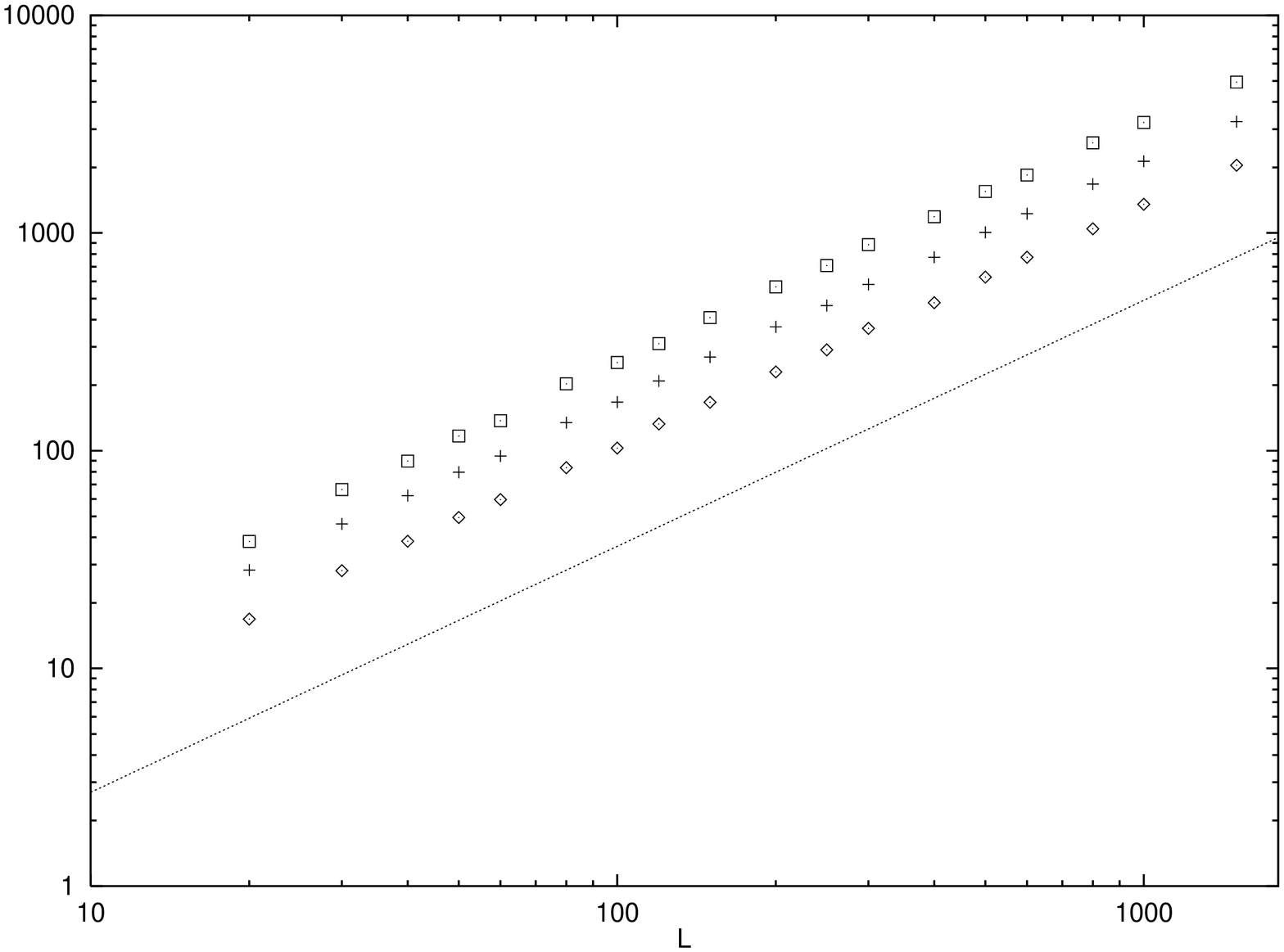, width = 3in, angle = 270}
\vskip 2 cm
\psfig {file = 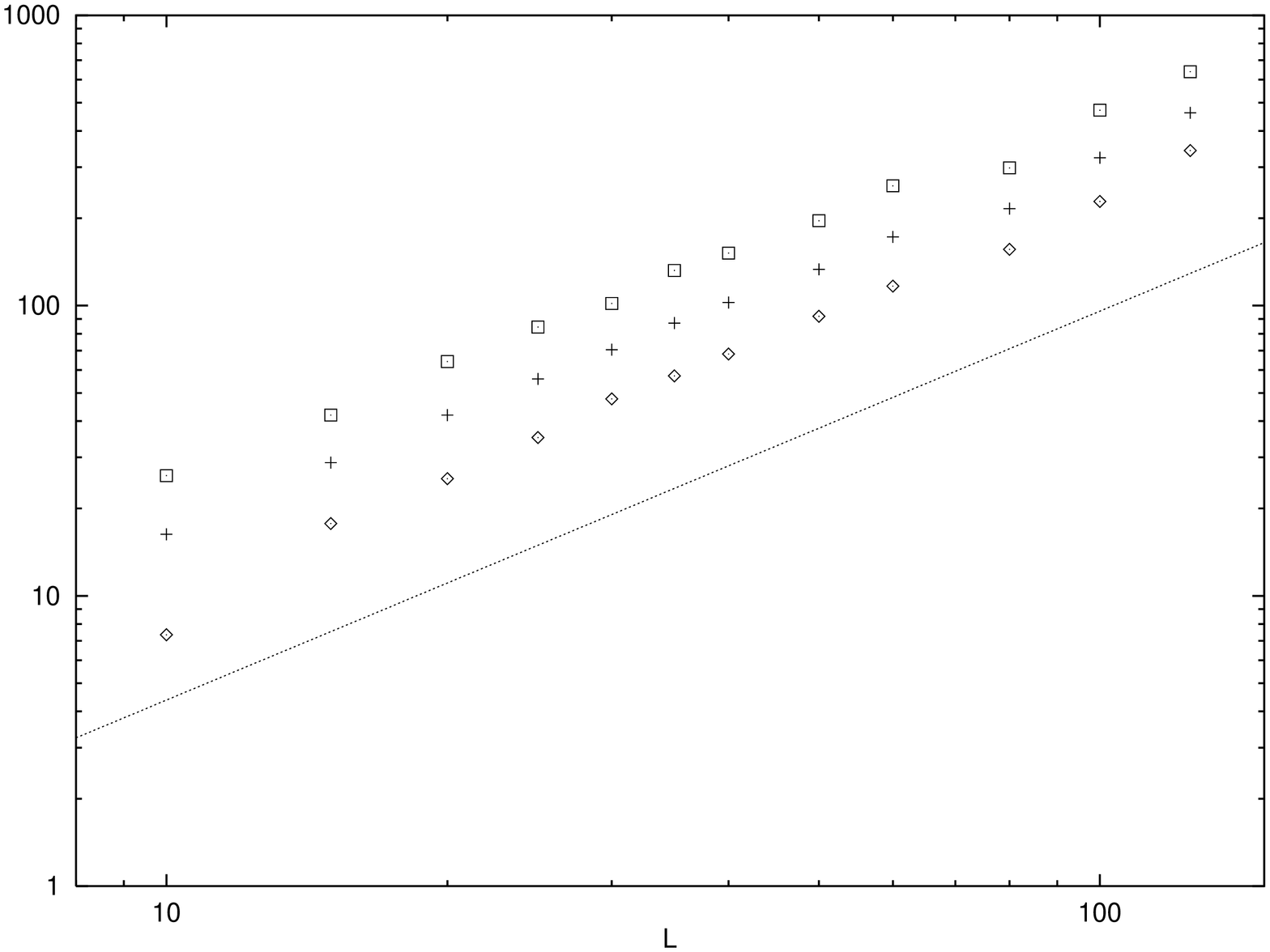, width = 3in, angle = 270}
\vskip 2 cm
\psfig {file = 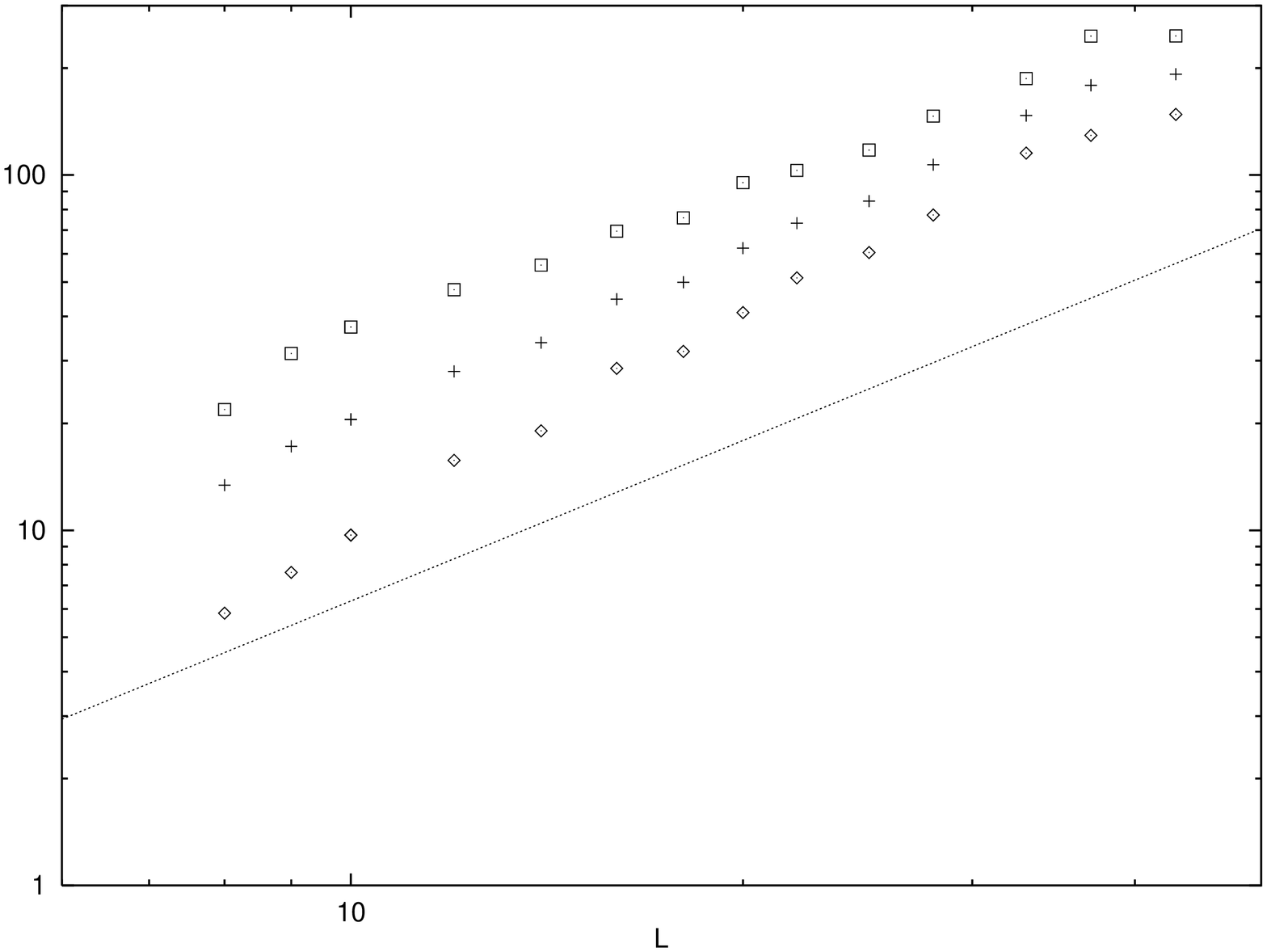, width = 3in, angle = 270}
\vskip 2 cm
\psfig {file = 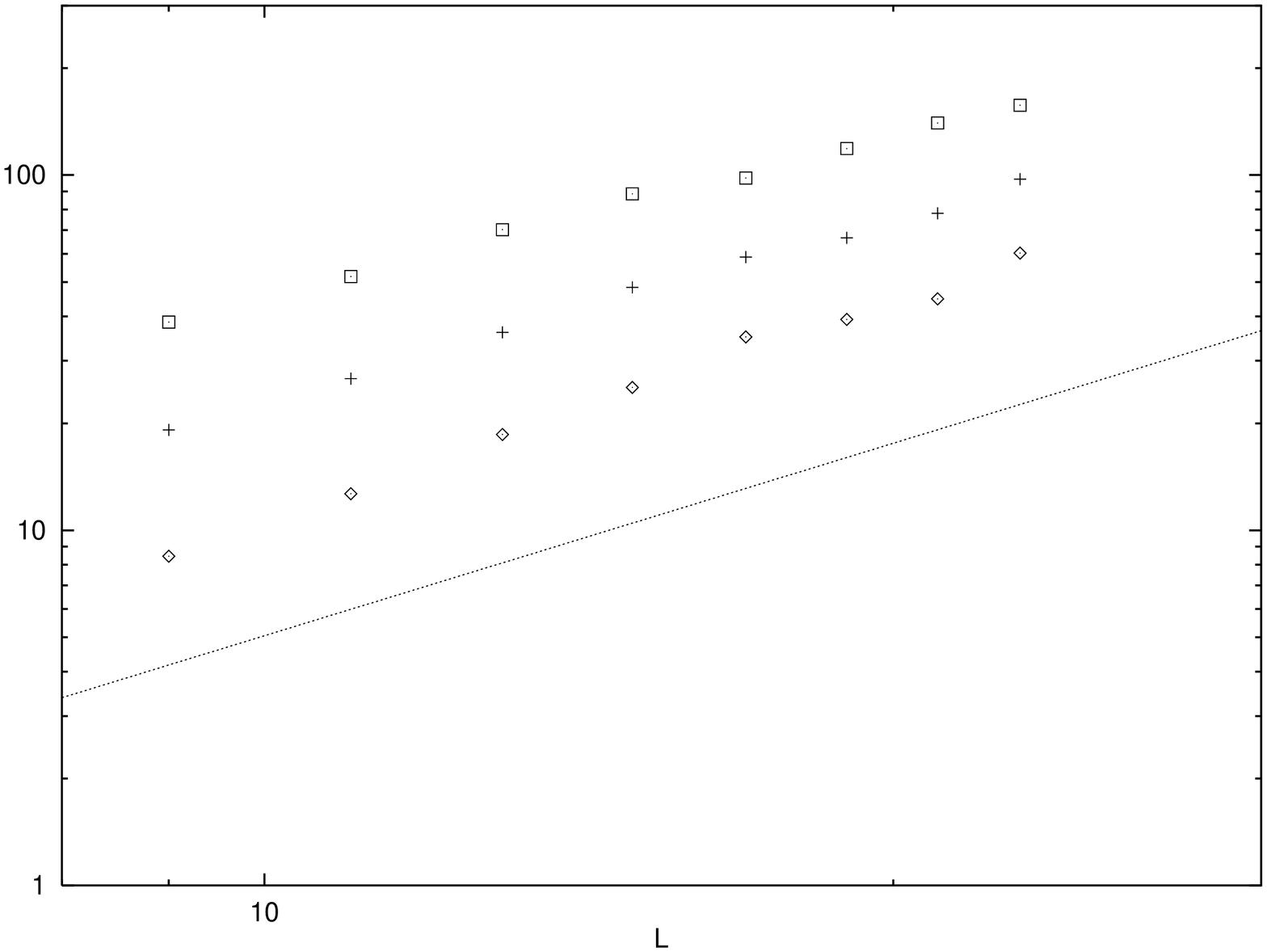, width = 3in, angle = 270}
\vskip 2 cm
\psfig {file = 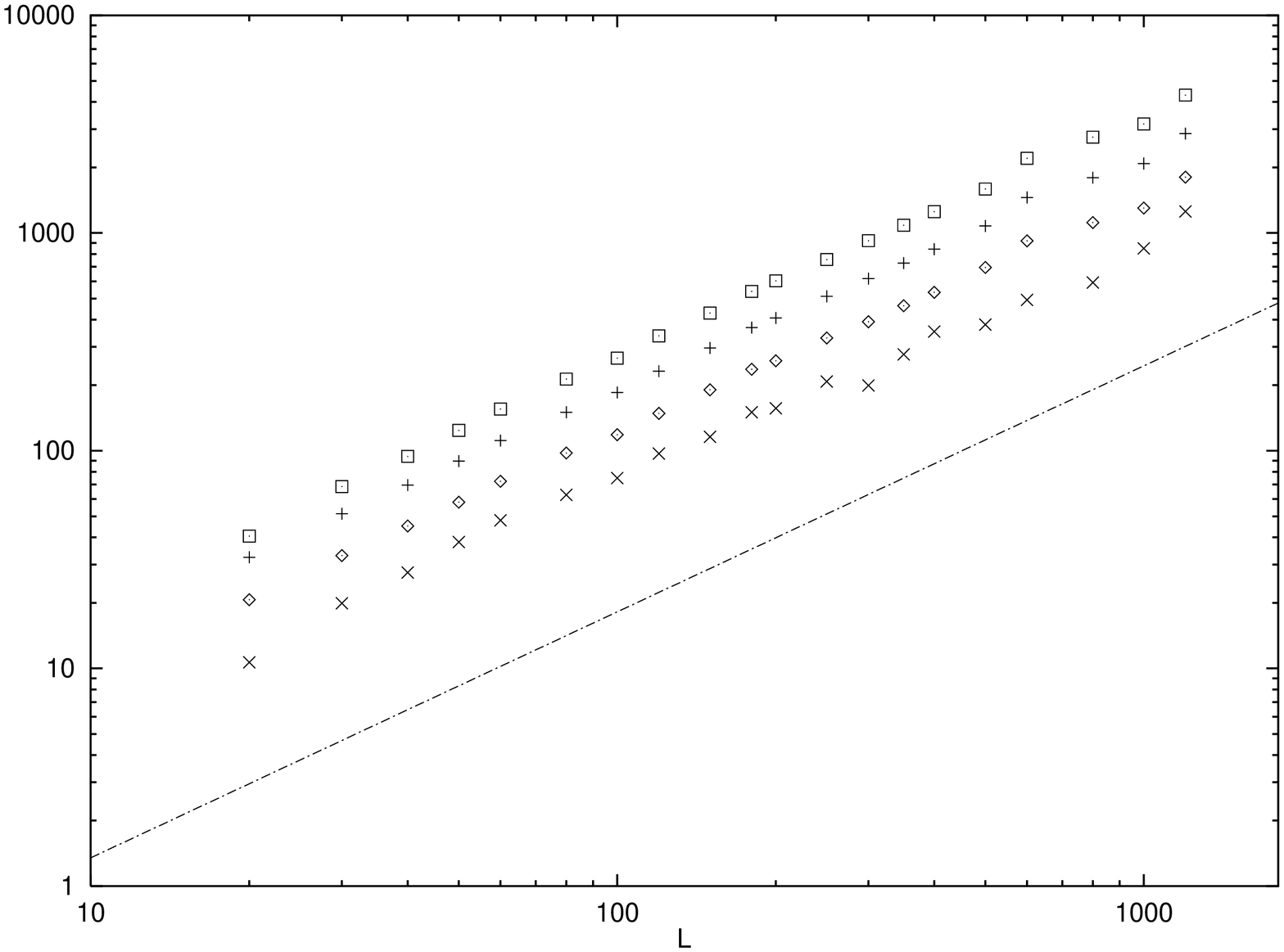, width = 3in, angle = 270}
\vskip 2 cm
\psfig {file = 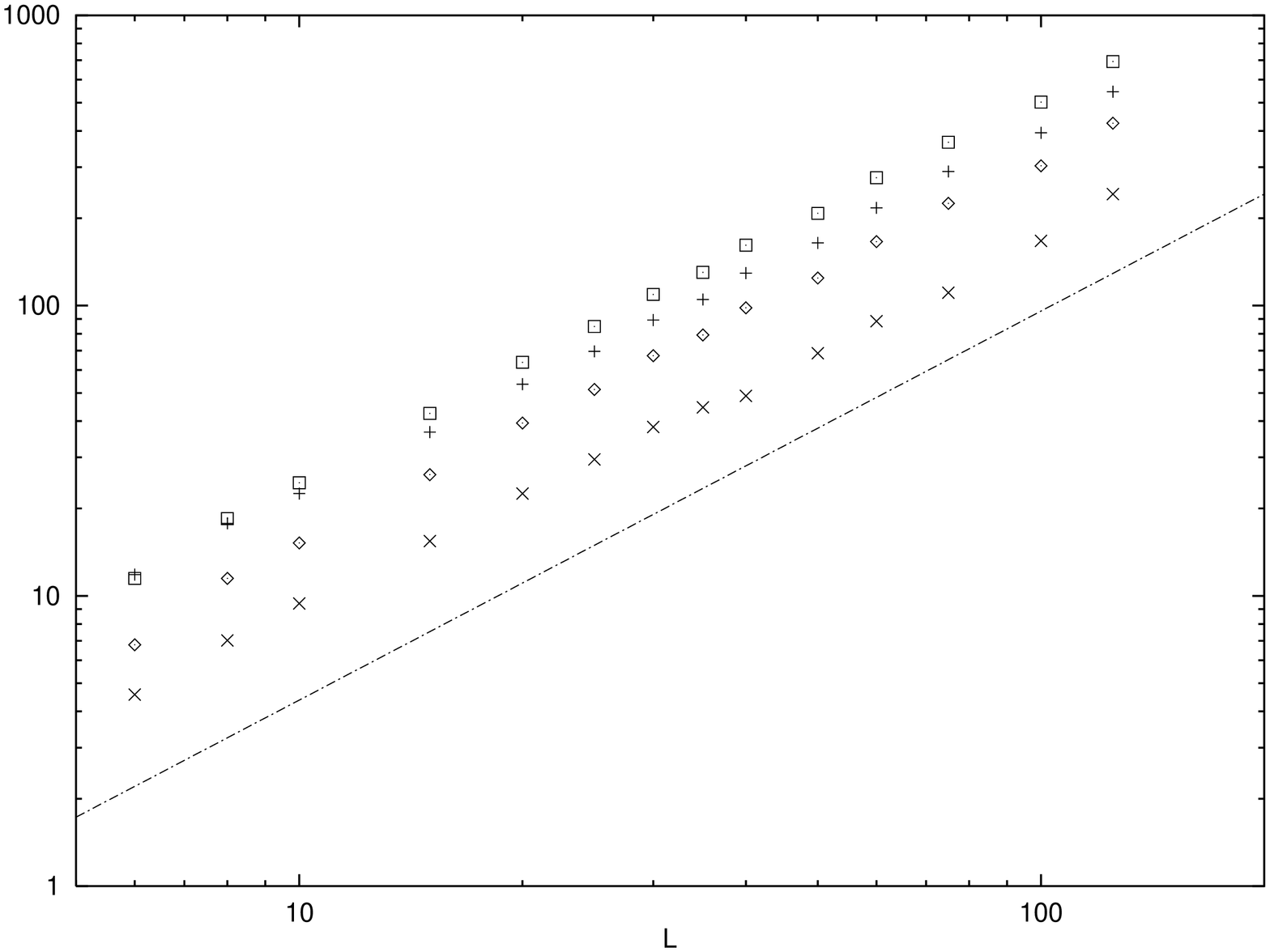, width = 3in, angle = 270}
\vskip 2 cm
\psfig {file = 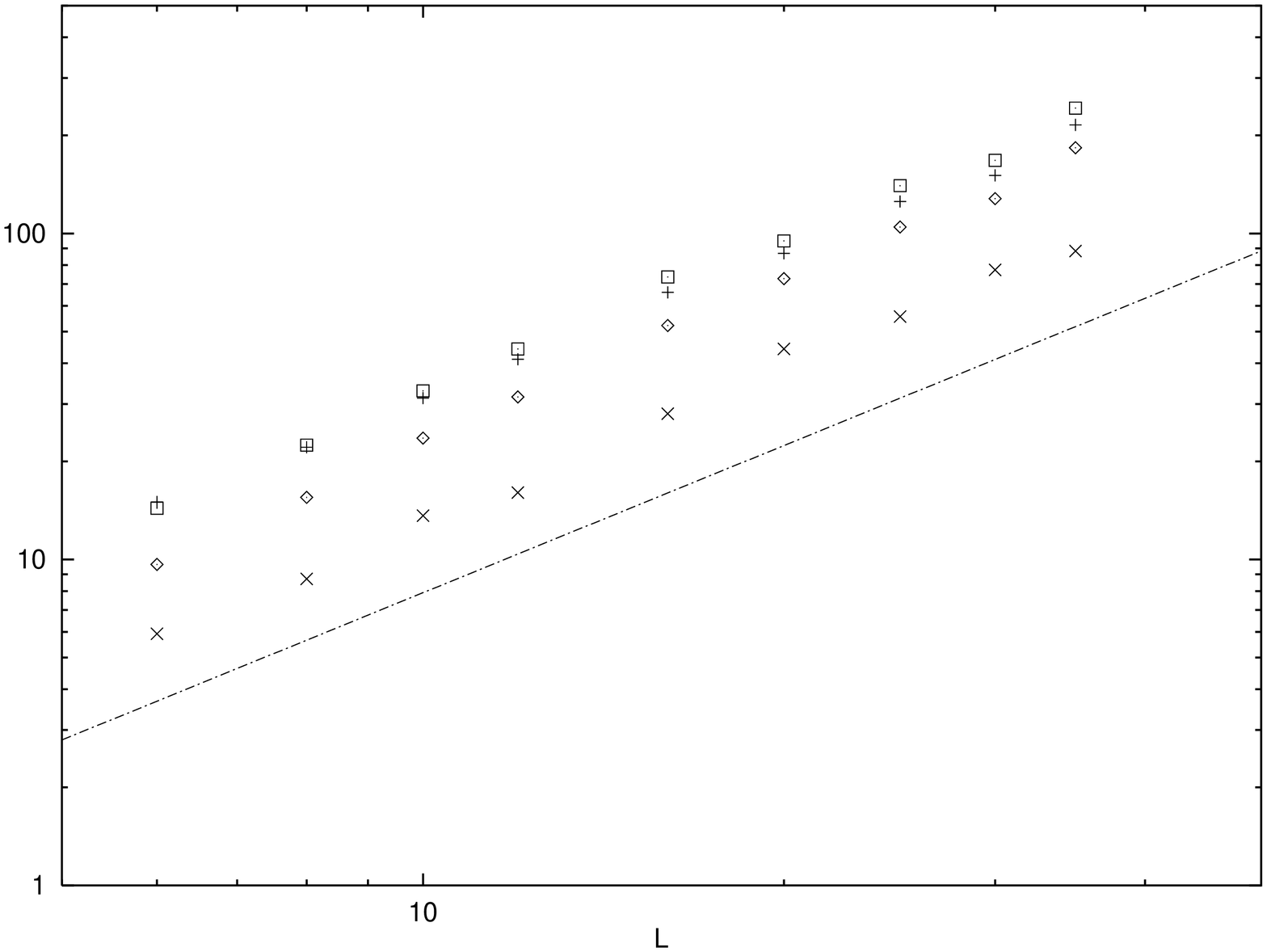, width = 3in, angle = 270}
\vskip 2 cm
\psfig {file = 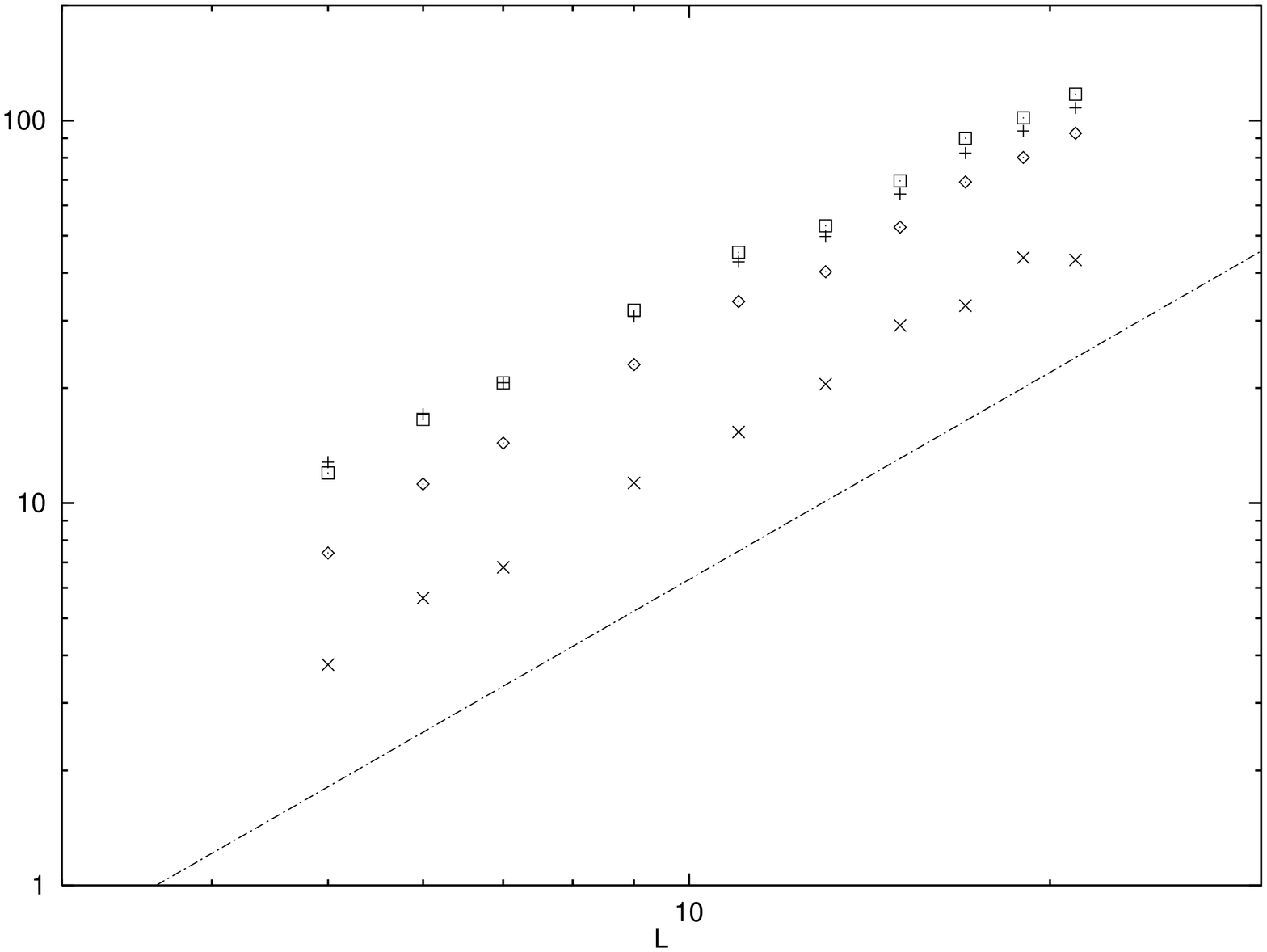, width = 3in, angle = 270}
\vskip 2 cm
\psfig {file = 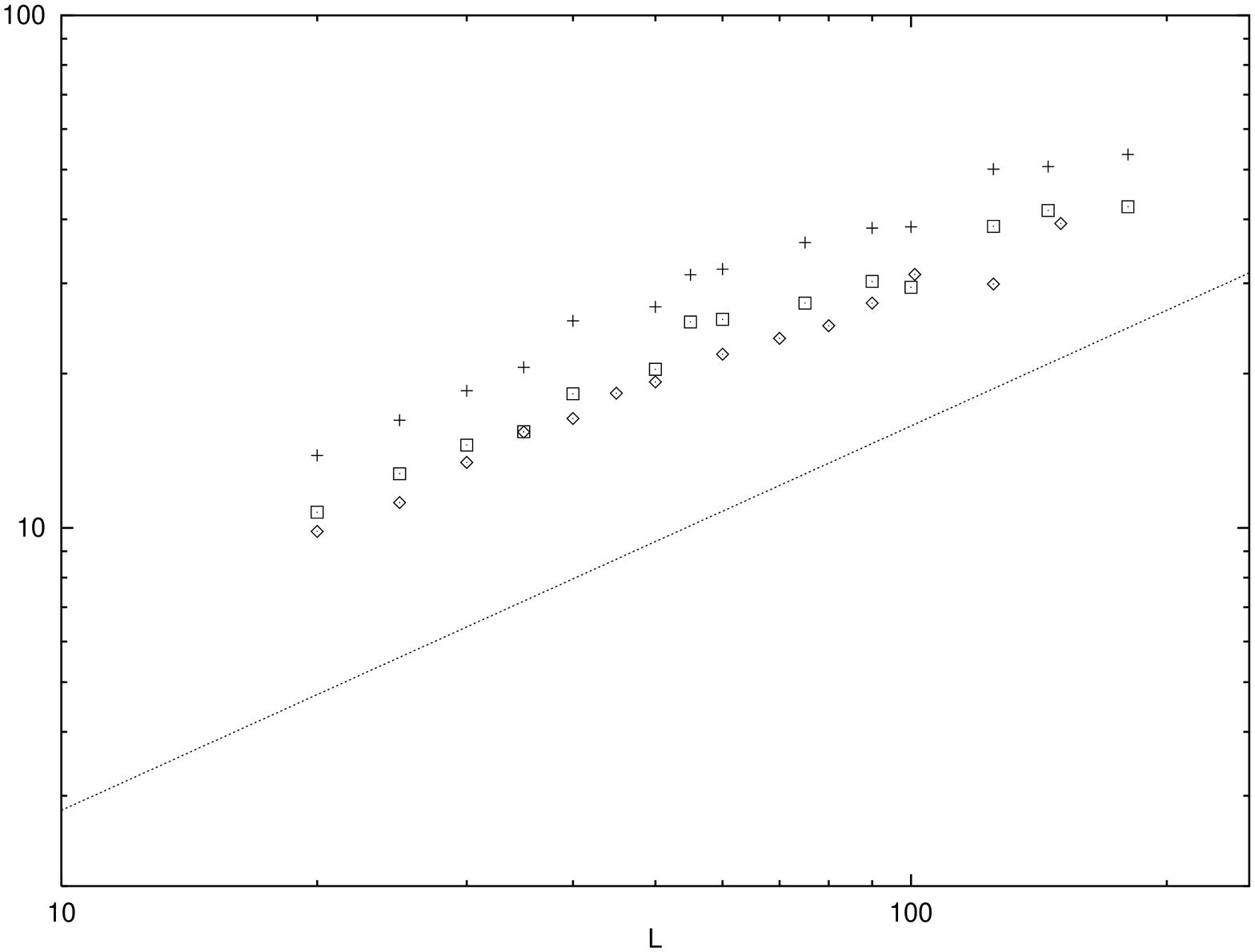, width = 3in, angle = 270}
\vskip 2 cm
\psfig {file = 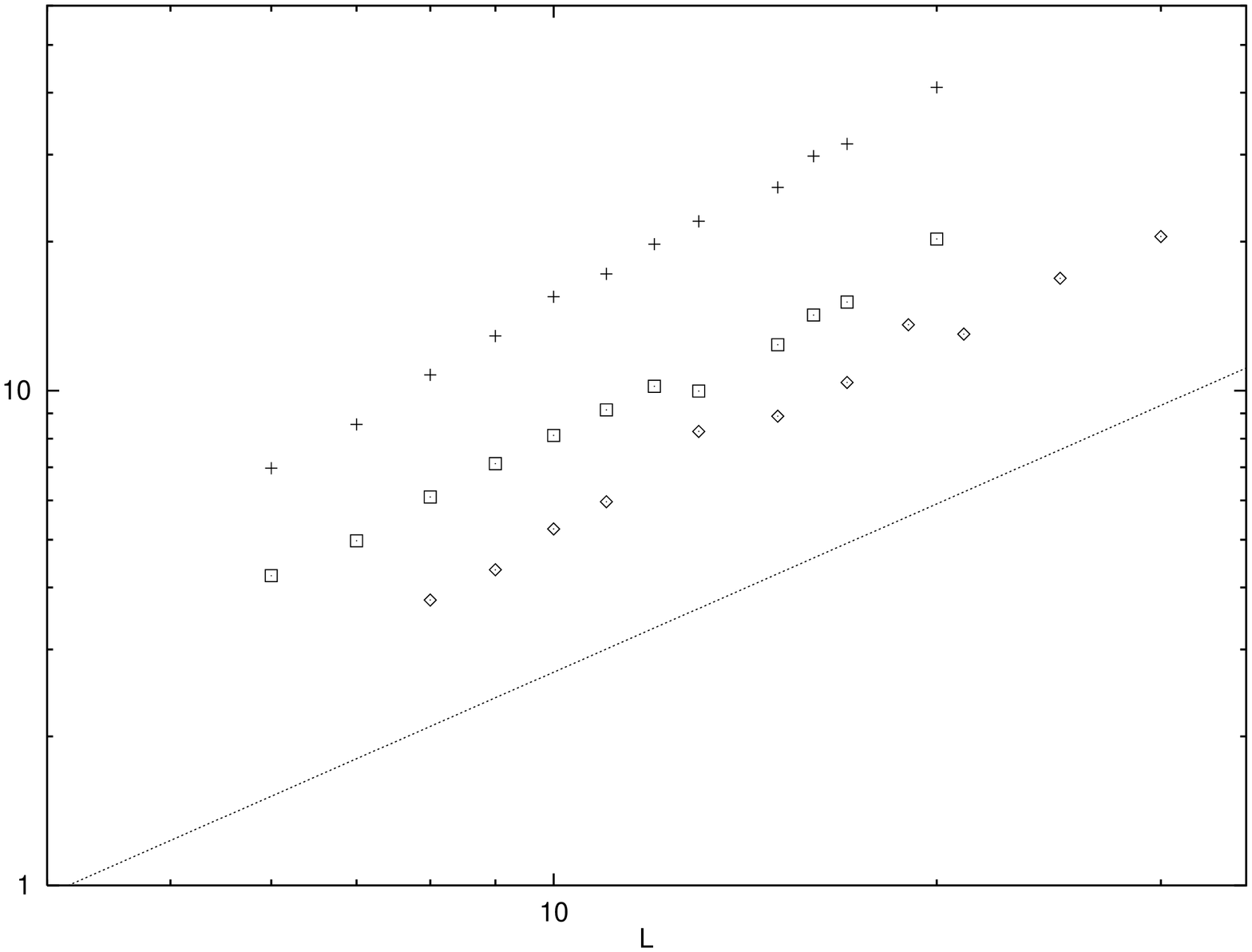, width = 3in, angle = 270}
\vskip 2 cm
\psfig {file = 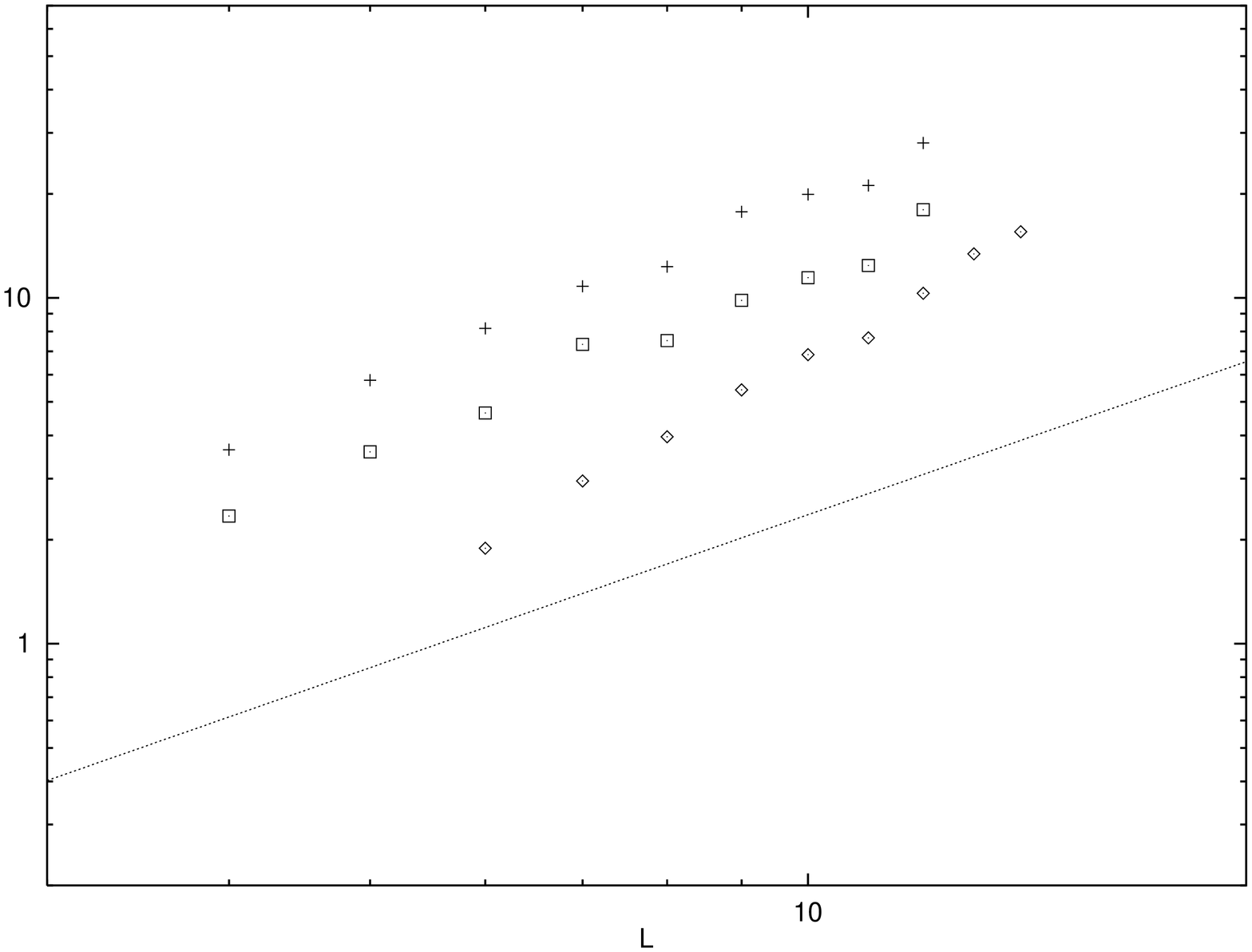, width = 3in, angle = 270}
\vskip 2 cm
\psfig {file = 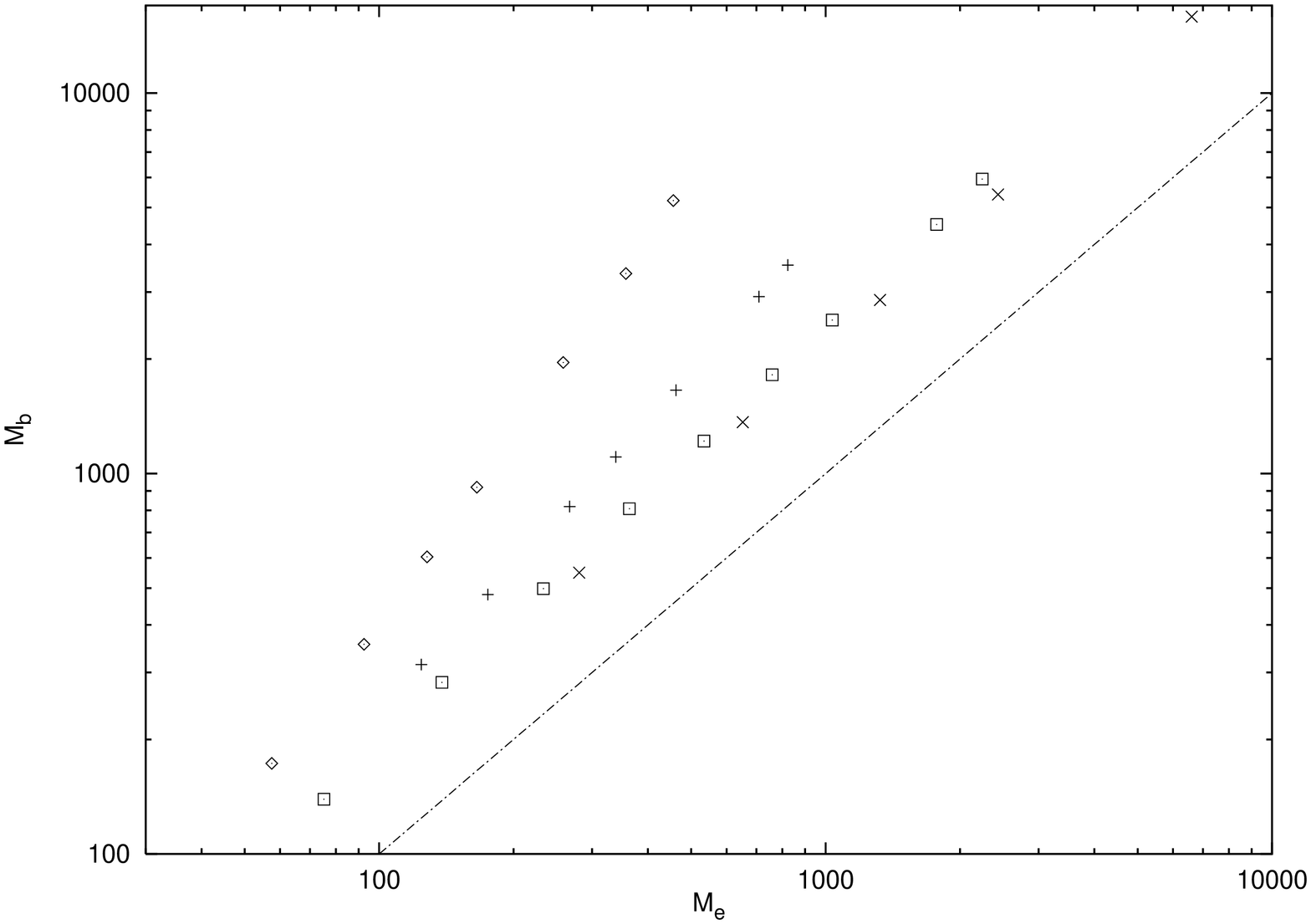, width = 3in, angle = 270}

\end{document}